\documentclass[]{llncs}
\usepackage{makeidx}
\usepackage{amssymb}
\usepackage{tikz,pgfplots}
\pgfplotsset{compat=newest}
\usetikzlibrary{positioning}
\usetikzlibrary{shapes.geometric,calc,matrix,arrows,decorations.pathmorphing}
\usetikzlibrary{3d,shapes,snakes}
\usetikzlibrary{shadings,mindmap,trees}
\tikzset{
	braces/.style = {
		outer sep=-1pt,
		left delimiter=(,
		right delimiter=),
		align=center,
	},
}

\newdimen\fwd
\newcommand{\N}{\mathbb{N}}
\newcommand{\R}{\mathbb{R}}
\newcommand{\SQN}{\ensuremath{\mathrm{S}q\mathrm{N}}}
\begin{document}

\newpage

\author{Kai~Brehmer\inst{1} \and
		Benjamin~Wacker\inst{1} \and
		Jan~Modersitzki\inst{1,2}}
\authorrunning{Kai~Brehmer et al.}
\tocauthor{Kai Brehmer, Benjamin Wacker, and Jan Moderistzki}
\title{A Novel Similarity Measure for Image Sequences}
\titlerunning{A Novel Similarity Measure}
\institute{Institute of Mathematics and Image Computing, 
University of L{\"u}beck, Germany, \\ \email{brehmer@mic.uni-luebeck.de} \and
Fraunhofer MEVIS, L{\"u}beck, Germany}

\maketitle

\begin{abstract}
	
Quantification of image similarity is a common problem in image processing. For
pairs of two images, a variety of options is available and
well-understood.
However, some applications such as dynamic imaging or serial sectioning
involve the analysis of image sequences and thus
require a simultaneous and unbiased comparison of many images.

This paper proposes a new similarity measure, that takes a global perspective and involves all images at the same time.
The key idea is to look at Schatten-$q$-norms of a matrix assembled from normalized gradient fields of the image sequence.
In particular, for $q=0$, the measure is minimized if the gradient information from the image sequence has a low rank.

This global perspective of the novel \SQN-measure does not only allow to register sequences from dynamic imaging, e.g. DCE-MRI,
but is also a new opportunity to simultaneously register serial sections, e.g. in histology.
In this way, an accumulation of small, local registration errors may be avoided.

First numerical experiments show very promising results for a DCE-MRI sequence of a human kidney as well as for a set of serial sections.
The global structure of the data used for registration with \SQN\ is preserved in all cases.

\end{abstract}

\section{Introduction}

Quantification of image similarity is a common problem in image processing. For
pairs of two images, a variety of options such as sum of squared differences (SSD), normalized gradient fields (NGF),
or mutual information (MI) exists and these measures are
well-understood; see e.g.~\cite{modersitzki04,modersitzki09,sotiras13}. 
However, some applications such as dynamic imaging or serial sectioning
involve the analysis of image sequences and thus
require a simultaneous and unbiased comparison of many images.

In some of these applications, image intensity may changes over time. 
For example, the glomerular filtration rate (GFR) is an important parameter for kidney malfunction~\cite{sourbron13}. The GFR might be determined on the basis  of a time series of dynamic contrast-enhanced magnetic resonance images (DCE-MRI). This sequence then needs to be registered in order to correct for motion artifacts; 
see, e.g., \cite{heck14,hodneland14}.
Another example is the analysis of a histological serial sectioning.
Here, the staining of sections might be different and/or can 
express severe variations.

For applications like these, a proper image similarity measure is crucial. A
standard approach is to perform a sequential comparison of pairs of two images
from the sequence. However, a sequential registration is restricted to local
rather than global information. Moreover, there might be issues choosing a
suited starting image and determining the order of the sequence.
Results may depend on these choices.
There also exists a variety of statistical approaches for global image registration; see e.g. \cite{guyader16,huizinga16,polfliet17,tao17}. However, we focus on a new global measure which is based on deterministic image features.

In this paper, we propose the new \SQN\ similarity measure, that is designed to
compare a complete image sequence simultaneously and thus automatically 
distributing information in a global way. Our key idea is to look at
Schatten-$q$-norms (more precisely: Schatten-$q$-quasinorms) of a matrix that is
assembled from normalized gradient fields of the image sequence. Particularly
for $q=0$, the measure is minimized for sequences of images where the gradient
matrix has low rank and the approach is thus connected to principal component analysis and sparsity.

Our idea is motivated from color image denoising; see
M\"{o}llenhoff et al.~\cite{moellenhoff15}. In that context, similar concepts are used
as a regularization for TV denoising, and the gradient matrix is formed
directly from gradients of the three color channels. In our paper, we interpret the individual
images from a sequence as individual channels and use a Schatten-$q$-norm of
the matrix of normalized gradients as a data fitting term rather than as a
regularizer. As we will show, this can be viewed as a natural extension of
NGF~\cite{haber05} and can thus deal with multi-modal frames. 
We remark that the concept also relates to ideas in video compression. 

\bigskip

Our paper is organized as follows:
At first we describe the novel distance measure and its relation to NGF as well as its characteristics. Moreover, we show numerical results for DCE-MRI time series and a H\&E stained histological serial section of a mouse brain.
We compare the performance of \SQN\ with NGF (DCE-MRI) and the well-known SSD (serial sectioning).
Our examples show that \SQN\ results at least comparable registration results but is about six times faster as the competitive approaches.
To conclude our paper, we discuss the numerical results and give a brief outlook on what our next steps are.


\begin{section}{The novel similarity measure \SQN}
	
Our new distance measure is motivated by a regularizer for color image
denoising; see~\cite{moellenhoff15}. The underlying idea is that in natural
images, gradients of the different color channels are linearly dependent;
see also Fig.~\ref{fig:colorchannels}. Therefore, an appropriate measure of
dependency such as Schatten-$q$-norms~\cite{watrous11} are excellent regularizers in color image denoising.
This idea can be generalized to more than three channels and is therefore
useful in applications such as parameter estimation in DCE-MRI~\cite{heck15}
or the registration of multiple images as they appear for example in serial
sectioning or time series \cite{lotz14}.

We motivate our extension starting with a conceptual simpler but computational
infeasible approach. The main point is to motivate the use of
Schatten-$q$-norms. We then present a computational tractable version that is
based on local image gradients. In contrast to~\cite{moellenhoff15} where a
similar measure is used as regularizer, we also propose to use our new functional as
a distance measure.

\begin{figure}[t]\centering

\begin{tikzpicture}[scale = 1.70]

	\begin{scope}
		\fill[gray!25]
		  (-3,-1) -- (-3,-0.8) -- (-2,0) -- (-1,-0.2) -- (-1,-1) -- cycle;
		\fill[gray!75]
		  (-3,-0.8) -- (-3,-0.2) -- (-2,0) -- cycle;
		\fill[black]
		  (-3,-0.2) -- (-3,0.2) -- (-2,0) -- cycle;
		\fill[gray!50]
		  (-2,0) -- (-1,0.2) -- (-1,-0.2) -- cycle;
		\fill[gray!60]
		  (-2,0) -- (-1,0.8) -- (-1,0.2) -- cycle;
		\draw (-3,-1) rectangle (-1,1);
		\draw[black,thick] (-3,0.2) -- (-1,-0.2);
		\draw[black,thick,->] (-2.0,0) -- (-1.85,0.75);
		\node[above] at (-1.85,0.75) {\textcolor{black}{$\nabla u_{3}$}};
		\draw[gray!25,thick] (-3,-0.8) -- (-1,0.8);
		\draw[gray!25,thick,->] (-2.0,0) -- (-2.4,0.5); 
		\node[above] at (-2.5,0.50) {\textcolor{gray!25}{$\nabla u_{2}$}};
		\draw[gray!75,thick] (-3,-0.2) -- (-1,0.2);
		\draw[gray!75,thick,->] (-2.0,0) -- (-2.1,0.5);
		\node[above] at (-2.1,0.50) {\textcolor{gray!75}{$\nabla u_{1}$}};
		\draw[fill=black](-2.0,0)circle(0.5pt);
	\end{scope}
	
	\begin{scope}
	    \fill[gray!50]
	      (-0.8,-0.5) -- (1.2,0.5) -- (1.2,-1) -- (-0.8,-1) -- cycle;
	    \fill[black]
	      (-0.8,-0.5) -- (1.2,0.5) -- (1.2,1) -- (-0.8,1) -- cycle;
		\draw (-0.8,-1) rectangle (1.2,1);
		\draw[fill=black](0.2,0.0)circle(0.5pt);
		\draw[gray!25,thick] (-0.8,-0.5) -- (1.2,0.5);
		\draw[gray!50,thick,->] (0.2,0.0) -- (-0.1,0.6);
		\node[above] at (-0.1,0.6) {\textcolor{gray!50}{$\nabla u_{1}$}};
		\draw[gray!25,thick,->] (0.2,0.0) -- (0.05,0.3);
		\node[right] at (0.05,0.3) {\textcolor{gray!10	}{$\nabla u_{2}$}}; 
		\draw[black,thick,->] (0.2,0.0) -- (0.4,-0.4);
		\node[below] at (0.4,-0.4) {\textcolor{black}{$\nabla u_{3}$}};
	\end{scope}
	
	\begin{scope}
		\draw (1.4,-1) rectangle (3.4,1);
		\fill[black]
		  (1.4,0.7) -- (3.4,-0.7) -- (3.4,-1) -- (1.4,-1) -- cycle;
		 \draw[gray!50,thick] (1.4,0.7) -- (3.4,-0.7);
		 \draw[fill=black](2.4,0.0)circle(0.5pt);
		 \draw[gray!25,thick,->] (2.4,0.0) -- (2.84,0.6);
		 \node[above] at (2.84,0.6) {\textcolor{gray!75}{$\nabla u_{2}$}};
	\end{scope}	
	
\end{tikzpicture}
	
\caption{Illustration of a local gradient matrix 
	$A=[\nabla u_{1},\nabla u_{2},\nabla u_{3}]\in \R^{2,3}$ of 
	three color channels; illustration adapted from~\cite{moellenhoff15}.
	The rank of $A$ is two (left) or one (center and right).}
\label{fig:colorchannels}
\end{figure}
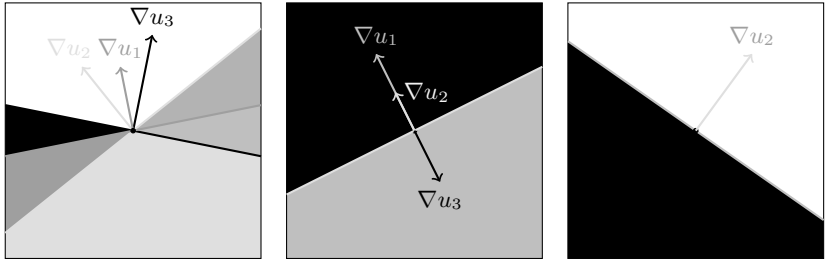

We recall that any matrix $A\in\R^{n,T}$ has a singular value decomposition (SVD)~\cite{golub12},
\[
	A=U\mathrm{diag}(S)V^{\top},
	\mbox{ with }
	U^{\top}U=E_n,\ 
	V^{\top}V=E_T.\ 
\]
Here, $E_d$ denotes the $d$-by-$d$ identity matrix,
$S=(\sigma_1,\ldots,\sigma_{\min\{n,T\}})$ is a vector with ordered entries
$\sigma_j\ge\sigma_{j+1}\ge0$ and $\mathrm{diag}(S)\in\R^{n,T}$ denotes a diagonal matrix. 
Using the SVD, the Schatten-$q$-(quasi)-norm of $A$ is then defined as
\[
	\|A\|_{S,q}^{q} 
	:=\|S\|_{q}^{q} 
	=\textstyle\sum_i\sigma_{i}^{q}
	\quad\mbox{for}\quad
	q \geq 0.
\]
An optimal choice of $q$ is obviously application dependent.
In particular for registration problems, it is topic of current research. Note that for $q=0$, the measure counts non-zero entries and is not a norm but a so-called quasinorm. With $q=0$, we thus promote sparsity of $S$ and hence low rank of $A$. However, optimization of the $0$-quasinorm is non-trivial and it is typically replaced by the minimization of the $1$-norm; see~\cite{candes08}.  
Following~\cite{moellenhoff15}, we use $q=1/2$ in this paper.

\bigskip

We are now ready to describe our novel similarity measure.
To this end, we assume that $T\in\N$ discrete images $I_t$ are given, where each $I_t$ is of dimension $m_1\times\cdots\times m_d\in\N^{d}$ 
and $d\in\N$ denotes the spatial dimension. 
Let $n:=m_1\cdots m_d$ and $I_t$ be reshaped as $n\times1$
array. 
The first idea is to look at the rank of $A=[I_1,\ldots,I_T]\in\R^{n,T}$ as an indicator for the linear dependency of the images. 
Note that this approach is similar to a principal component analysis of the data.
However, this approach is not suitable for images with varying intensities such as DCE-MRI or serial sections.
Although this approach is conceptually appealing, it is computationally challenging as the complexity of the SVD is 
$\mathcal{O} \left(\min \{ nT^2, Tn^2 \} \right)$~\cite{golub12}.

We escape the complexity problem by applying the measure to local structures.
More precisely, we define our new distance measure \SQN\ for 
a sequence of $T$ images by
\begin{eqnarray*}
	\SQN(I_{1},\ldots,I_{T})
	&:=& \int \|A(x)\|_{S,q}^{q}\ \mathrm{d}x,	
	\\
	\noalign{where}
	A(x)&:=&[\eta_1,\ldots,\eta_T]\in\R^{d,T}
	\quad\mbox{and}\\
	\eta_t&:=&(\nabla I_t(x)^{\top}\nabla I_t(x)+\eta^2)^{-1/2}\ \nabla I_t(x).
\end{eqnarray*}
Here, $\eta>0$ is a parameter discriminating signal from noise; see also~\cite{haber05}. If the noise level is unknown, we pick a small value, e.g. $\eta=10^{-5}$ in our experiments where every entry of $I_t$ lies in $[0,256)$.

We now outline the connection to NGF~\cite{haber05}.
For two images $I_1$ and $I_2$, fixed $x$, $\eta=0$ and with
$\alpha:=|\cos\angle(\nabla I_1,\nabla I_2)|$, we have
\[
	A^{\top}A 
	=\Big(\begin{array}{rr} 
      1 & \alpha \\	\alpha & 1
	  \end{array}\Big)
	  \mbox{ with singular values }
	  \sigma_{1}^2=1+\alpha
	  \mbox{ and }
	  \sigma_{2}^2=1-\alpha.
\]
Particularly for $q=0$, the distance is minimal if the image gradients are collinear, i.e. $\alpha=1$. Therefore, our new measure might be interpreted as a generalization of the normalized gradient field based distance measure~\cite{haber05}. Analogous arguments hold for $0\le q<2$. 
Remarkably, for $q=2$, the energy function is constant and the measure thus meaningless. For $q>2$, the function is minimal for perpendicular gradients.

\end{section}

\begin{section}{Numerical Results for Dynamic Imaging and Serial Sections}

We now present results for the registration of DCE-MRI sequences of a human
kidney and a histological serial sectioning of a mouse brain.

\bigskip

We start with registrations of DCE-MRI sequences of a human
kidney; data courtesy of Jarle R{\o}rvik, Haukeland University Hospital Bergen, Norway.
Here, 3D images are taken at 49 time
points. The objective is to register the images while maintaining the dynamics.
More precisely, we use 146-by-82 coronal slices of a 146-by-82-by-52-by-49 volume for z-Slice 25; see
Fig.~\ref{fig:results}. 

Our registration scheme is based on the variational image registration framework FAIR~\cite{modersitzki09}. More precisely, we minimize
\[ 
  \mathcal{J}(y_1,\dots,y_T) 
  =\SQN(I_1(y_1),\dots,I_T(y_T)) 
  +\textstyle\alpha\sum_{t=1}^T\ \mathcal{S}(y_t).
\] 
For ease of presentation, we use the curvature regularizer
$\mathcal{S}(y)=\int(\Delta y)^2\ dx$ with regularization parameter
$\alpha=0.1$~\cite{fischer03}. 
Optimization is performed in a standard way using a
Gau{\ss}-Newton algorithm with Armijo linesearch~\cite{nocedal06} within a
multi-level framework~\cite{modersitzki09}. All computations are performed
using MATLAB.

\fwd=35mm
\newbox\myImg
\setbox\myImg=\hbox{\includegraphics[height=\fwd,angle=90]{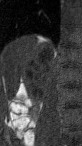}}	
\begin{figure}[t]\centering
	\begin{tabular}{ccc}
		data-$t_1$ & data-$t_2$ & data-$t_3$
		\\
		\includegraphics[height=\fwd,angle=90]{./fig/img1-1}
	&   \includegraphics[height=\fwd,angle=90]{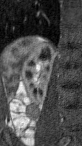}
	&   \includegraphics[height=\fwd,angle=90]{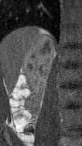}
	\\
	   MIP-original & MIP-\SQN & MIP-NGF
	   \\
  	   \includegraphics[width=\fwd]{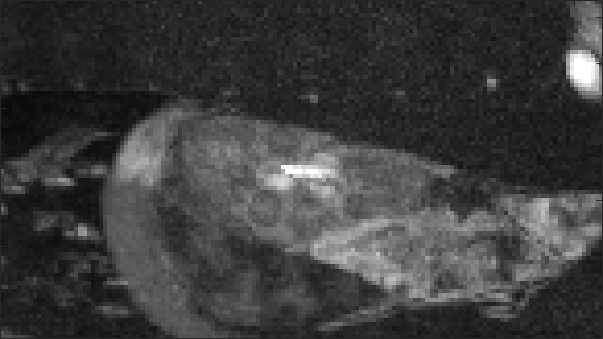}
	  &\includegraphics[width=\fwd]{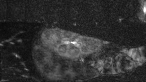}
	  &\includegraphics[width=\fwd]{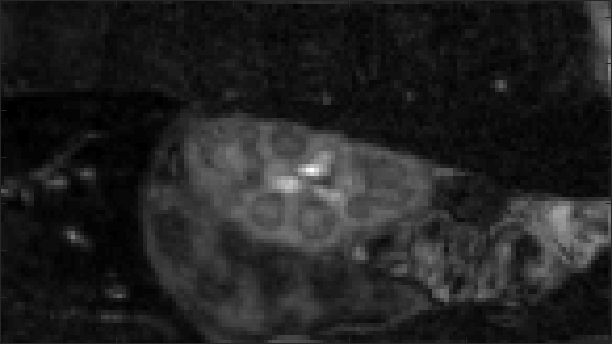}
	\end{tabular}

\caption{%
	DCE-MRI data of a human kidney;
	data courtesy of Jarle R{\o}rvik, Haukeland University Hospital Bergen, Norway.
	Top row: Displayed are 2D slices at three representative time points
	during contrast agent uptake.
	Images are rotated by 90 degrees for presentation purposes.
	Bottom row: Coronal view of maximum intensity projections
	$\sum_{j>i}|I_j-I_i|$ for original, 
	\SQN-registered, and NGF-registered data.
	Note the blurred and doubled structures in the non-registered data.
	}
\label{fig:results}
\end{figure}

\fwd=40mm
\begin{figure}
  \begin{tabular}{ccc@{\quad}cc}
	 & sagittal-1
	 & sagittal-2
	 & axial-1
	 & axial-2
	 \\
	 \rotatebox{90}{original}
    &\includegraphics[height=\wd\myImg,angle=90]{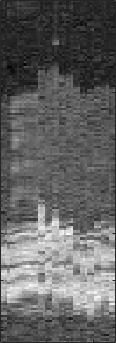}
    &\includegraphics[height=\wd\myImg,angle=90]{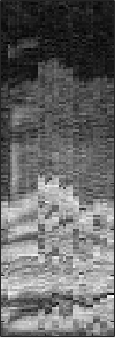}
    &\includegraphics[height=\ht\myImg,angle=90]{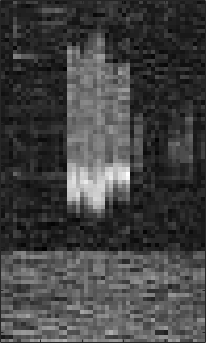}
    &\includegraphics[height=\ht\myImg,angle=90]{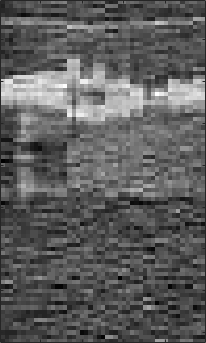}
    \\
	 \rotatebox{90}{\hskip2mm\SQN}
    &\includegraphics[height=\wd\myImg,angle=90]{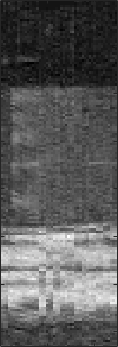}
    &\includegraphics[height=\wd\myImg,angle=90]{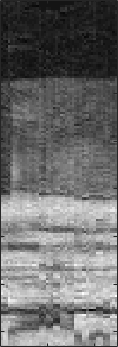}
    &\includegraphics[height=\ht\myImg,angle=90]{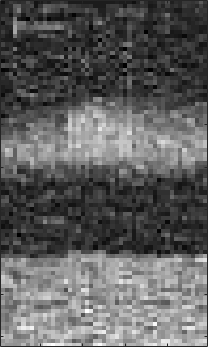}
    &\includegraphics[height=\ht\myImg,angle=90]{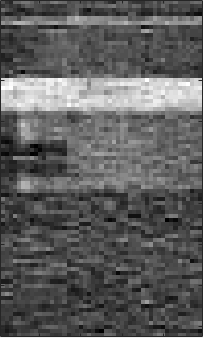}
	\\
	 \rotatebox{90}{\hskip2mm NGF}
		&\includegraphics[height=\wd\myImg,angle=90]{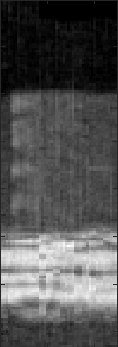}
		&\includegraphics[height=\wd\myImg,angle=90]{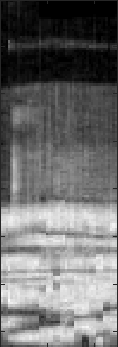}
		&\includegraphics[height=\ht\myImg,angle=90]{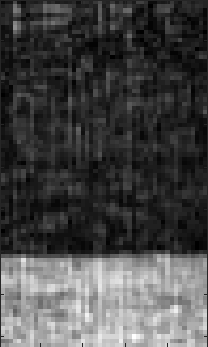}
		&\includegraphics[height=\ht\myImg,angle=90]{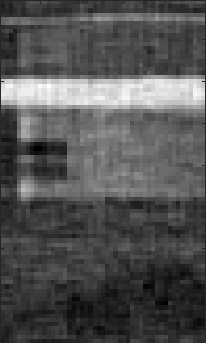}
  \end{tabular}
  
\caption{%
  Two exemplary sagittal and axial slices of the data, each;
  see also Fig.~\ref{fig:results}.
  Original, non-registered data (top row),
  \SQN-registered data (middle row, $q=0.5$), and
  NGF-registered data (bottom row), $\eta=25$.
  Note that the laminar structure of the tissue is only visible
  after registration. The axial sections in this visualization
  do not necessarily correspond.
}
  \label{fig:results_sag}
\end{figure}

Fig.~\ref{fig:results} (top row) shows three representative coronal slices of
the original dataset. The slices correspond to different times during
acquisition. Different phases of contrast agent uptake are visible,
particularly within the kidney.

Fig.~\ref{fig:results_sag} displays sagittal and axial slices of the same volume, two each. The top row shows non-registered slices where
motion artefacts are clearly visible.  
The middle row shows corresponding \SQN-registration results for $q=0.5$
and the bottom row shows results for sequential NGF, respectively.
More precisely, we optimized
\[
  \mathcal{J}^{\mathrm{NGF}}(y_1,\dots,y_T) 
  =\textstyle\sum_{t=1}^{T-1}\Big\{\ 
   \mathrm{NGF}(I_{t}(y_{t}),I_{t+1}(y_{t+1}))+\alpha\ \mathcal{S}(y_t)
  \Big\}
\] 
using alternating optimization, i.e.
\[
	y_{t}^{k+1}
	=\mathrm{argmin}_{y_t}\ \mathcal{J}^{\mathrm{NGF}}(
	 y_1^{k+1},\dots,y_{t-1}^{k+1},\ y_t,\ y_{t+1}^{k},\ldots,y_T^{k}),
	 \quad
	 t=1,\ldots,T.
\]

As to be expected, the \SQN\ and NGF results are very similar.
However, within our non-optimized MATLAB framework, 
the \SQN-registration is about six times faster than the NGF
registration --- even if the alternating optimization approach is limited to a forward-backward sweep.

Fig.~\ref{fig:results} also illustrates the intensity variations
in the original and registered data. It is apparent that intensity variations due to motion have been reduced tremendously.
Note that the variations, in particular in kidney and bladder (partly visible), are still visible and therefore the schemes allow for a subsequent dynamical analysis.

\bigskip

The \SQN\ measure is also capable of serial section registration.
Fig.~\ref{fig:results_HNSP} displays results for a H\&E stained 
histological serial section of a mouse brain;
189 sections of 512-by-512 pixel;
data courtesy of O.~Schmitt, University of Rostock, Germany;
see \cite{schmitt01} for experimental details.
As this data has been normalized, we compare an \SQN\ registration with a sequential SSD based approach.
The figure displays the original data (top row), \SQN\ results (middle row),
and SSD results (bottom row).
It is apparent that registration can reconstruct the local brain structure.
Here, the SSD results appear to be slightly more blurred.

Note that a sequential approach involves an alternating optimization framework
and is based on a fixed initial and final slice to avoid the so-called
banana-effect~\cite{gaffling11,schmitt01,streicher97,wang2015}, i.e. a global drift of structures due to sequential registraton. 
The sequential registration process may accumulate small errors that can cause
a major drift of the overall structure; see~\cite{wang2015} for examples. 

In contrast, the \SQN\ approach enables a global optimization which in addition can be performed in just one pass. Moreover, our experiments indicate
that \SQN\ does not introduce global drifts.

\bigskip

Fig.~\ref{fig:energyComp} shows the results of a translation experiment for the distance measures SSD, MI, NGF as wells as for \SQN\ for three different configurations of the parameter $q$.
It is apparent that $q$ has an impact on the global minimum of the energy in this experiment. All measures have the same minimizer which is achieved for the template image in its origin when it is not shifted as shown in the difference image of Fig.~\ref{fig:energyComp}.

\fwd=35mm
\newbox\myImg
\setbox\myImg=\hbox{\includegraphics[height=\fwd,angle=180]{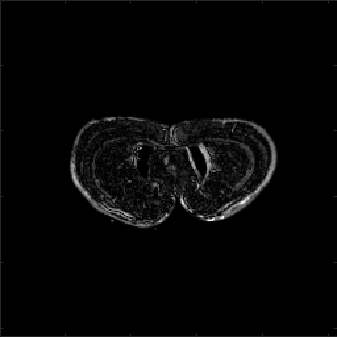}}	
\begin{figure}[t]\centering
	\begin{tabular}{ccc}
		\includegraphics[height=\fwd,angle=180]{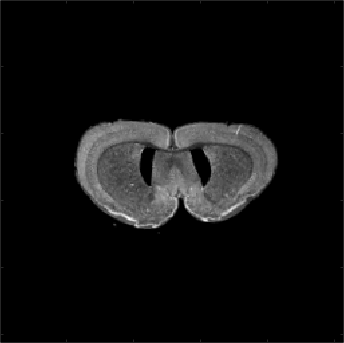}
		&   \includegraphics[height=\fwd,angle=180]{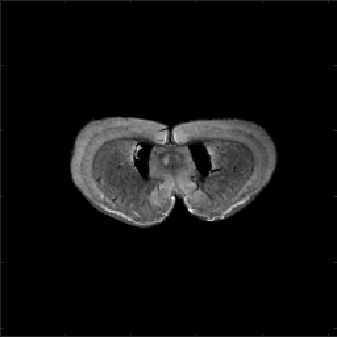}
		&   \includegraphics[height=\fwd,angle=180]{./fig/energyCompDiff}
		\\
		Reference & Template & Absolute Difference
	\end{tabular}

	\vspace{2em}
	\resizebox{2.5\fwd}{!}{\input{./fig/energyComparisonPlots.tex}}
	
	\caption{%
		Results for a translation of two stained histological serial sections; 
		data courtesy of O.~Schmitt, University of Rostock, Germany~\cite{schmitt01}.
		The top row displays the similar slices 160 (reference) and 170 (template) of a stack 
		of 189 slices in total as well as the absolute difference image.
		The bottom row displays the corresponding energies of the different distance measures listed
		in the legend. The images used for translation are of size $256 \times 256$ pixel.
		The y axis of the graphs are scaled individually for better comparison.
	}
	\label{fig:energyComp}
\end{figure}

\begin{figure}[t]\centering\fwd=35mm
	\begin{tabular}{cccc}
		& axial & coronal & sagittal
		\\
		\rotatebox{90}{\hskip10mm original}
		&\includegraphics[height=\fwd,angle=90,origin=center]%
		{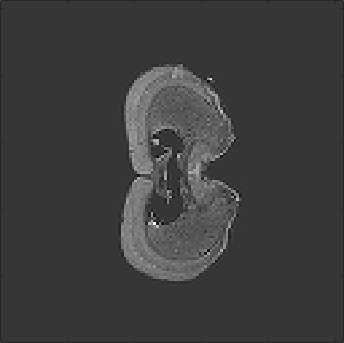}
		&\includegraphics[height=\fwd]%
		{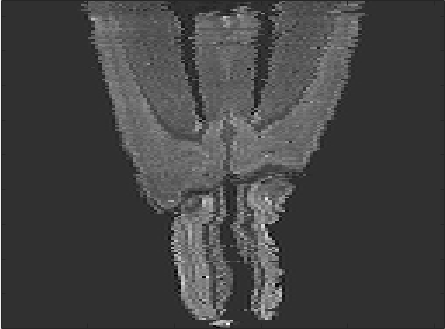}
		&\includegraphics[height=\fwd]%
		{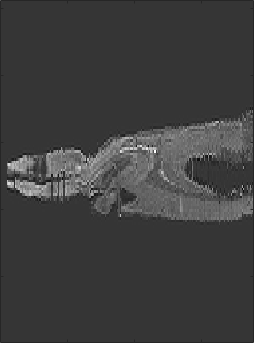}
		\\
		\rotatebox{90}{\hskip10mm \SQN, $q=0.5$}
		&\includegraphics[height=\fwd,angle=90,origin=center]%
		{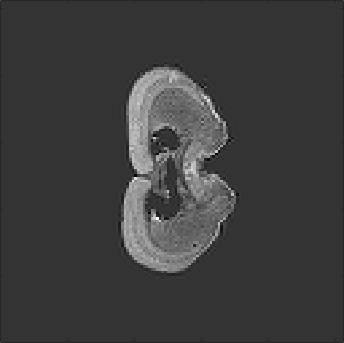}
		&\includegraphics[height=\fwd]%
		{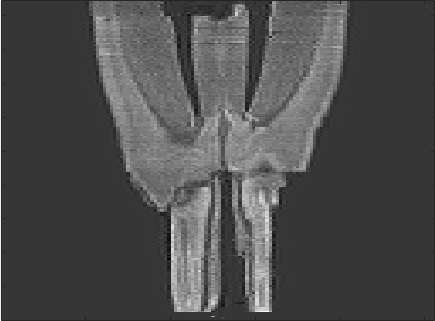}
		&\includegraphics[height=\fwd]%
		{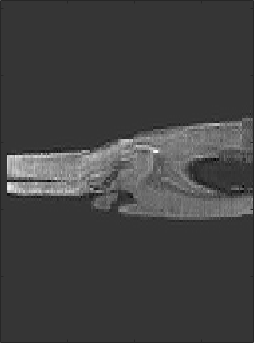}
		\\
		\rotatebox{90}{\hskip10mm SSD}
		&\includegraphics[height=\fwd,angle=90,origin=center]%
		{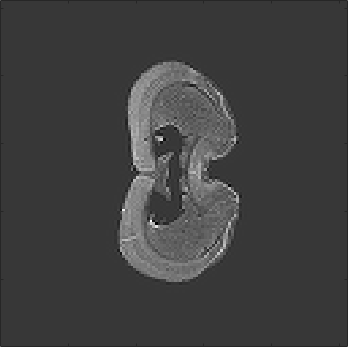}
		&\includegraphics[height=\fwd]%
		{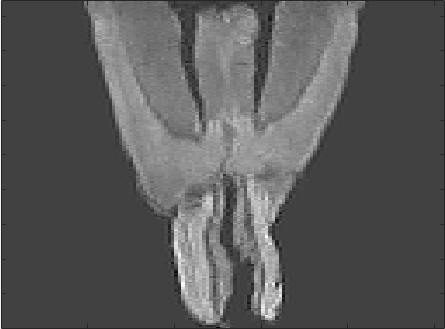}
		&\includegraphics[height=\fwd]%
		{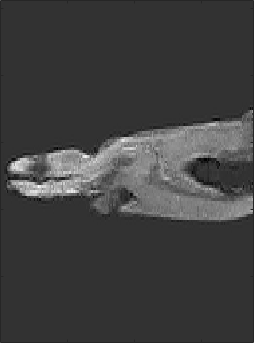}
		\\
	\end{tabular}
	
	\caption{%
		Registration results for a stained histological serial sectioning;
		data courtesy of O.~Schmitt, University of Rostock, Germany~\cite{schmitt01}.
		Displayed from left to right are exemplarily
		an axial, coronal, and sagittal slice of the 3D data of size
		512-by-512-by-189.\newline
		Displayed are non-registered data (top row),
		\SQN-registered data (middle row) and 
		SSD-registered data (bottom row).
		Note that the different slices do not necessarily correspond.
	}
	\label{fig:results_HNSP}
\end{figure}

\end{section}

\section{Discussion and Conclusions}

The novel \SQN\ image similarity measure has been proposed. The new measure is
motivated from color image denoising \cite{moellenhoff15}. The main idea is to quantify
structural image information expressed by normalized intensity gradients with
Schatten-q-(quasi)-norms. For $q=0$, the measure quantifies sparsity, i.e.
redundancy of information in an image sequence. Moreover, using the normalized
gradient fields as structural information, the focus is on image structures and
not on intensity. Therefore, the new measure is effective in a multi-modal setup and it might be interpreted as an extension of NGF~\cite{haber05}.

The novel measure considers the image sequence as a whole and therefore has no
bias towards a particular ordering, which is common in a sequential setup.
Therefore, the new measure provides also a global information transport, which might be beneficial for particular applications.
In particular, it omits an unwanted drift of structure as it is very common in sequential approaches; so-called banana-effect~\cite{streicher97}.

We show the potential of the new measure in a registration setup. More
precisely, we show how easily the new measure can be integrated in an existing
registration framework such as FAIR~\cite{modersitzki09}. Both, the measure and its analytic
derivative are computed and thus, efficient optimization techniques can be used.

Exemplarily, we demonstrate the power of the measure in a registration problem for a dynamic contrast enhanced MRI sequence and histological serial sectioning. As it turns out, the new scheme produces results of at least comparable quality much faster (about six times faster in our experiments). 

Particularly for serial section registrations, we escape an additional iterative process on the sequences of sections and make use
of global information transport.

Our next steps include to come up with a more efficient implementation which is suitable in a 3D setup. We remark that the main computational cost is not the
computation of the SVD of the gradient matrix. This involves the computation of eigenvalues of a $T$-by-$T$ matrix, where $T$ is the number of images in the sequence. In particular, $T$ does not dependent on the spatial dimension of the data or the spatial resolution. However, the computation of the gradient matrix $A$ does.

Moreover, we will investigate the impact of the distance measure in the analysis of the dynamic information within the DCE-MRI setting. With this knowledge, we would be able to quantify an optimal choice for the parameter $q$. In our current experiments, it appears 
that the impact of $q$ is rather small unless we pick values close to $q=2$. We also will compare the measure to global statistical measures (e.g. \cite{guyader16,huizinga16,polfliet17,tao17}) already used in this field of registration and image processing.

An implementation of \SQN\ will soon be available within the FAIR-Toolbox; see https://github.com/C4IR.

\section*{\small Acknowledgement}
\begin{minipage}{0.5\textwidth}
	\centering
	\includegraphics[scale = 0.6,trim = 0.8cm 1.3cm 1.5cm 0.5cm]{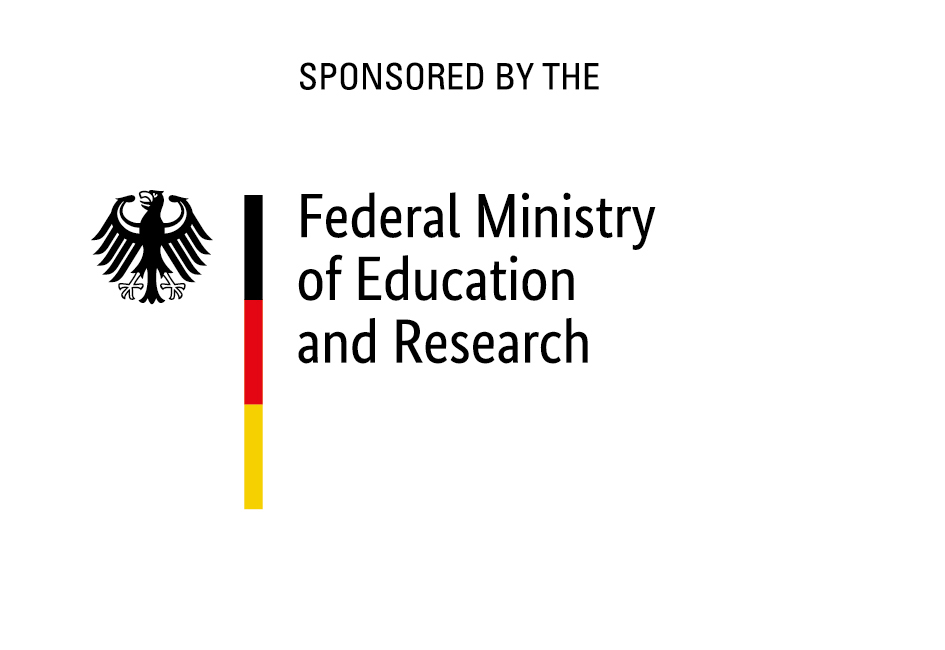}
\end{minipage}
\begin{minipage}{0.5\textwidth}
	The authors acknowledge the financial support by the Federal Ministry of Edu\-cation and Research of Germany in the framework of MED4D (project number 05M16FLA).
\end{minipage}


\end{document}